\documentclass[reprint,superscriptaddress,preprint,preprintnumbers,bibnotes, amsmath,amssymb, aps]{revtex4}
\usepackage{graphicx}
\usepackage{epstopdf}
\usepackage{color}
\usepackage{xcolor}
\usepackage{soul}
\usepackage{ulem}
\usepackage{dcolumn}
\usepackage{float}
\usepackage{hyperref}
\hypersetup{hypertex=true,
            colorlinks=true,
            linkcolor=blue,
            anchorcolor=blue,
            citecolor=blue}
\usepackage{bm}
\begin{document}
\preprint{}
\title{Raman Spectroscopic Study on Bi$_2$Rh$_3$Se$_2$: Two-dimensional-Ising Charge Density Wave and Quantum Fluctuations}

\author{Fei Jiao}
\affiliation{Anhui Key Laboratory of Condensed Matter Physics at Extreme Conditions, High Magnetic Field Laboratory,Chinese Academy of Sciences, Hefei 230031, China}
\affiliation{Science Island Branch of Graduate School, University of Science and Technology of China, Hefei 230026, China}
\author{Yonghui Zhou}
\affiliation{Anhui Key Laboratory of Condensed Matter Physics at Extreme Conditions, High Magnetic Field Laboratory,Chinese Academy of Sciences, Hefei 230031, China}
\affiliation{Science Island Branch of Graduate School, University of Science and Technology of China, Hefei 230026, China}
\author{Shuyang Wang}
\affiliation{Anhui Key Laboratory of Condensed Matter Physics at Extreme Conditions, High Magnetic Field Laboratory,Chinese Academy of Sciences, Hefei 230031, China}
\author{Chao An}
\affiliation{Institutes of Physical Science and Information Technology, Anhui University, Hefei 230601, China}
\author{Xuliang Chen}
\affiliation{Anhui Key Laboratory of Condensed Matter Physics at Extreme Conditions, High Magnetic Field Laboratory,Chinese Academy of Sciences, Hefei 230031, China}
\affiliation{Science Island Branch of Graduate School, University of Science and Technology of China, Hefei 230026, China}
\author{Ying Zhou}
\affiliation{Institutes of Physical Science and Information Technology, Anhui University, Hefei 230601, China}
\author{Min Zhang}
\affiliation{Institutes of Physical Science and Information Technology, Anhui University, Hefei 230601, China}
\author{Liang Cao}
\affiliation{Anhui Key Laboratory of Condensed Matter Physics at Extreme Conditions, High Magnetic Field Laboratory,Chinese Academy of Sciences, Hefei 230031, China}
\author{Xigang Luo}
\affiliation{Department of Physics, University of Science and Technology of China, and Key Laboratory of Strongly Coupled Quantum Matter Physics, Chinese Academy of Sciences, Hefei 230026, China}
\author{Yimin Xiong\footnote[1]{Email: yxiong@ahu.edu.cn}}
\affiliation{Department of Physics, School of Physics and Optoelectronics Engineering, Anhui University, Hefei 230601, China}
\affiliation{Hefei National Laboratory, Hefei 230028, China}
\author{Zhaorong Yang\footnote[2]{Email: zryang@issp.ac.cn}}
\affiliation{Anhui Key Laboratory of Condensed Matter Physics at Extreme Conditions, High Magnetic Field Laboratory,Chinese Academy of Sciences, Hefei 230031, China}
\affiliation{Science Island Branch of Graduate School, University of Science and Technology of China, Hefei 230026, China}
\affiliation{Institutes of Physical Science and Information Technology, Anhui University, Hefei 230601, China}

\date{\today}

\begin{abstract}
The ternary chalcogenide Bi$_2$Rh$_3$Se$_2$ was found to be a charge density wave (CDW) superconductor with a 2$\times$2 periodicity.The key questions regarding the underlying mechanism of CDW state and its interplay with lattice and electronic properties remains to be explored.Here, based on the systematic Raman scattering investigations on single crystalline Bi$_2$Rh$_3$Se$_2$, we observed the fingerprinting feature of CDW state, a collective amplitude mode at $\sim$ 39 $cm^{-1}$. The temperature evolution of Raman shift and line width for this amplitude mode can be well described by the critical behavior of two-dimensional (2D) Ising model, suggesting the interlayer interactions of Bi$_2$Rh$_3$Se$_2$ is negligible when CDW state is formed, as a consequence, the quantum fluctuations play an important role at low temperature. Moreover, temperature dependence of Raman shift for  A$_{\rm g}^9$ mode deviates significantly from the expected anharmonic behavior when approaching the CDW transition temperature $\sim$  240 K, demonstrated that strong electron-phonon coupling plays a key role in the formation of CDW. Our results reveal that Bi$_2$Rh$_3$Se$_2$ is an intriguing quasi-2D system to explore electronic quantum phase transition and modulate the correlations between CDW and superconductivity.

\end{abstract}

\maketitle

\section{\label{sec:level1}INTRODUCTION}
Charge density wave(CDW) is of great interest due to its intertwined connections with multiple quantum states of matter, such as superconductivity (SC) and antiferromagnetism (AFM)\cite{OrtizBR,JiangY,TengX}, as well as the intriguing mechanisms of forming ordered charge density distribution\cite{ZhuXT}. CDW usually occurs in low-dimensional systems and involves periodic modulation of electron density, and this electronic modulation is accompanied by lattice distortion driven by a soft phonon mode, serving as a distinct fingerprint of the CDW order\cite{Weber}. Hence, it is of particular interest to clarify the nature of the soft mode in CDW states: the behavior of the soft mode, which refers to the phonon mode with a frequency approaching zero, is one of the most fundamental and extensively studied phenomena associated with classical (thermally driven) displacive phase transitions\cite{GrunerG}. When a soft phonon mode condenses and causes lattice distortions, a new Raman-active collective excitation, called amplitude mode, emerges, offering valuable insights into the evolution and stability of the CDW states\cite{RiceMJ,GrunerG,SugaiS}. Meanwhile, in low dimensional systems, the quantum fluctuation effects are more pronounced, leading to numerous unexpected and exotic properties with the interplay of strong correlations, electron-phonon coupling and other quantum effects\cite{yugate,ugeda,stojchevska}.

Recently, the compound Bi$_2$Rh$_3$Se$_2$ has attracted attentions as a CDW material with the coexistence of superconductivity(SC). Sakamoto et al. first synthesized Bi$_2$Rh$_3$Se$_2$ polycrystalline samples and revealed a CDW transition at $T \sim$ 240 K\cite{SakamotoT}. Later Chen et al. found symmetry-forbidden reflections for C12/m1 in selected-area electron diffraction (SAED) patterns and anomalous pressure dependence of the phase transition temperature, suggesting a pure structural phase transition accompanied by symmetry reduction(from C12/m1 to P12/m1) instead of a CDW transition in Bi$_2$Rh$_3$Se$_2$\cite{ChenCY}. Recently, experiments of Angle-resolved photoemission spectroscopy (ARPES) and infrared spectroscopy presented new evidences of Bi$_2$Rh$_3$Se$_2$ as a CDW metal with a 2 $\times$ 2 modulation of the crystal lattice\cite{LinT,LiuZT}. More recently, Wang et al. demonstrated the existence of incommensurate and commensurate CDW orders below 250 K and 170 K in Bi$_2$Rh$_3$Se$_2$\cite{WangYS}. Despite these experimental results, the thermodynamic property of CDW state at low temperature, along with the correlations to SC of Bi$_2$Rh$_3$Se$_2$ have not been explored. The amplitude mode of Raman spectra would provide an idea pathway to understand the forming mechanism of CDW as well as its characteristics.

In this work, by Angle-Resolved polarized Raman measurement, a multitude of CDW-induced modes were observed and studied on Bi$_2$Rh$_3$Se$_2$ single crystals. Temperature dependence of amplitude mode ($\sim$ 39 cm$^{-1}$) is consistent with the critical behavior of two-dimensional Ising model. The fast increase in the Raman shift and line width of amplitude mode exhibit the evidences for strong quantum fluctuations. The Raman shift of phonon-mode is clearly deviates from the expected anharmonic model when approaching \textit{T}$_{\rm CDW}$, indicating that electron-phonon coupling plays an important role on the formation of CDW phase\cite{WangSY}. The simultaneous presence of two-dimensionality, strong quantum fluctuation and electron-phonon coupling provides the clues for understanding the coexisting CDW and SC states in Bi$_2$Rh$_3$Se$_2$.

\section{\label{sec:level1}EXPERIMENT}
High quality single crystals of Bi$_2$Rh$_3$Se$_2$ were prepared by the method reported in the literature\cite{LinT}. Temperature-dependent resistivity was measured by using the standard four-probe method in a Quantum Design Physical Property Measurement System(PPMS) equipped with a $^3$He refrigerator. Magnetic susceptibility from 1.8 to 300 K was measured under the field of 1 Tesla by using a superconducting quantum interference device magnetometer(SQUID). A freshly exfoliated single crystal with an exposed surface of \textit{ab}-plane was transferred to a JANIS-ST500 cryogenic refrigerator at a temperature of 4.2 - 300 K and a vacuum better than 1$\times$10$^{-6}$ mbar for Raman measurements. Raman scattering spectra were collected using a Renishaw inVia microscope system with backscattering with a 532 nm laser. The laser beam is focused on the sample with a 20 $\times$ objective, and the laser power is kept at 0.5 mW to reduce the laser heating effect and protect the sample from damage. Single-crystal and powder X-ray diffraction experiments were performed with a PANalytical X$^{\prime}$pert diffractometer using Cu K$\alpha$1 radiation ($\lambda$ = 0.15406 nm).
\section{RESULTS AND DISCUSION}

\begin{figure*}
\includegraphics[width=16cm]{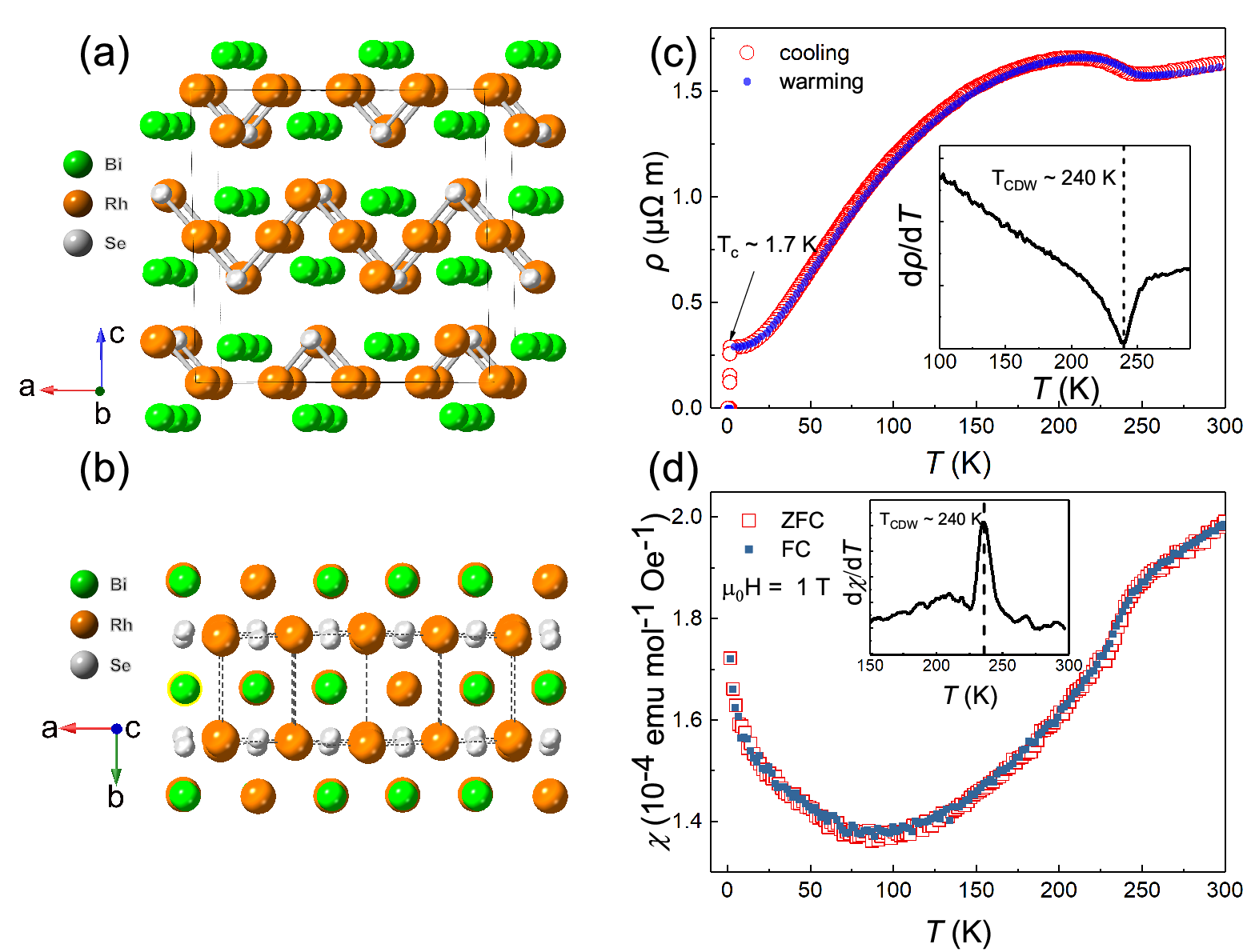}
\caption{Physical properties of Bi$_2$Rh$_3$Se$_2$. The schematic crystal structure of Bi$_2$Rh$_3$Se$_2$ are shown along the \textit{b}-axis (a) and \textit{c}-axis (b), respectively. (c) Temperature dependence of in-plane electrical resistivity $\rho$ of Bi$_2$Rh$_3$Se$_2$, the arrow indicates the temperature of \textit{T}$_{\rm c}$. 
The inset shows the derivative of resistivity d$\rho$/dT, with dashed line marking the CDW transition temperatures. (d) Temperature dependence of magnetic susceptibility measured at magnetic field $\mu_0$H = 1 Tesla with both zero-field-cooling (ZFC) and field-cooling (FC) process. The inset shows the derivative of susceptibility d$\chi$/dT, with black dashed line marking the CDW transition temperatures.}
\end{figure*}

Bi$_2$Rh$_3$Se$_2$ has layered parkerite-type ternary chalcogenides structure stacking along the \textit{c}-axis, as displayed by the schematic figures of crystal structure in Fig. 1(a). As shown in Fig. 1(b), the quasi-two-dimensional sheets contain rectangular network of Rh with short interatomic Rh-Rh distances 2.86 {\AA} along $a$-axis and the longer one along $b$-axis\cite{SakamotoT}. At \emph{T} $\textgreater$ $T_{CDW}$, the Rh networks could be treated as quasi-one-dimensional (1D) chains along the \textit{a}-axis due to its shorter distance. The data of powder XRD diffractions (at 300 K and 35 K) and analysis by standard Rietveld refinement method are plotted in Fig. S1(a) of Supplementary Materials. The fitting parameters are summarized in Table S1, which are consistent with the previously reported values\cite{SakamotoT}. The single crystal XRD pattern is displayed in Fig. S1(b). Only (00\emph{l}) diffraction peaks can be detected, indicating that the exposed surface of single crystals is $ab$-plane.

Fig. 1(c) shows temperature dependence of the in-plane resistivity of a Bi$_2$Rh$_3$Se$_2$ single crystal. There is a broad hump-like resistivity anomaly below $\sim$ 240 K, corresponding to the CDW transition reported in previous reports\cite{SakamotoT,LinT,LiuZT,WangYS}. The metallic behavior of resistivity confirms the partial gap opening at the Fermi surface in the CDW state\cite{SinghaR}. No thermal hysteresis can be observed in the resistivity curves between the cooling and heating processes, indicating that this CDW phase transition is a second-order one. The inset of Fig. 1(c) shows the derivative of resistivity d$\rho$/dT with an extremum at 240 K. A superconducting transition at $\textit{T}_{\rm c}$  $\sim$ 1.7 K is observed, which is higher than the reported value in polycrystal samples\cite{SakamotoT}, indicating that the SC in Bi$_2$Rh$_3$Se$_2$ is strongly affected by disorder\cite{LiJ,mayohanisotropic}.

Fig. 1(d) shows the magnetic susceptibility as a function of temperature (1.8 - 300 K). A knee-like change of the slope for both field-cooling (FC) and zero-field-cooling (ZFC) susceptibility curves appears near the CDW transition temperature. As shown in the inset of Fig. 1(d), the derivative of susceptibility also has an extremum at $T$ $\sim$ 240 K and no hysteresis was observed, in agreement with the resistivity data and results of the previous report\cite{SakamotoT}. Moreover, the susceptibility curves present a minimum value at around 100 K. These changes of susceptibility with temperature may be related to the change of the Fermi surface and density of states caused by the formation of CDW states. The Hall resistivity was analysed to understand the evolution of Fermi surfaces and density of states with temperature, as shown in Fig. S2 of Supplementary Materials. Around T$_{CDW}$ $\sim$ 240 K, the carrier density shows a slight temperature dependence, suggesting the anomalies in resistivity and susceptibility are not caused by the change of carrier density. At 240 K, the reconstruction of Fermi surfaces due to the formation of CDW state may induce a partial gapping of the Fermi surface below T$_{CDW}$ and orbital contribution to susceptibility, resulting in a hump-like anomaly in resistivity and a drop in $\chi$(T). Furthermore, the carrier density increases rapidly in commensurate CDW state and reaches a maximum value at about 100 K, which is consistent to the feature of the susceptibility curve in Fig. 1(d), indicating a dramatic change of Fermi surfaces and density of states.

\begin{figure*}
\includegraphics[width=14cm]{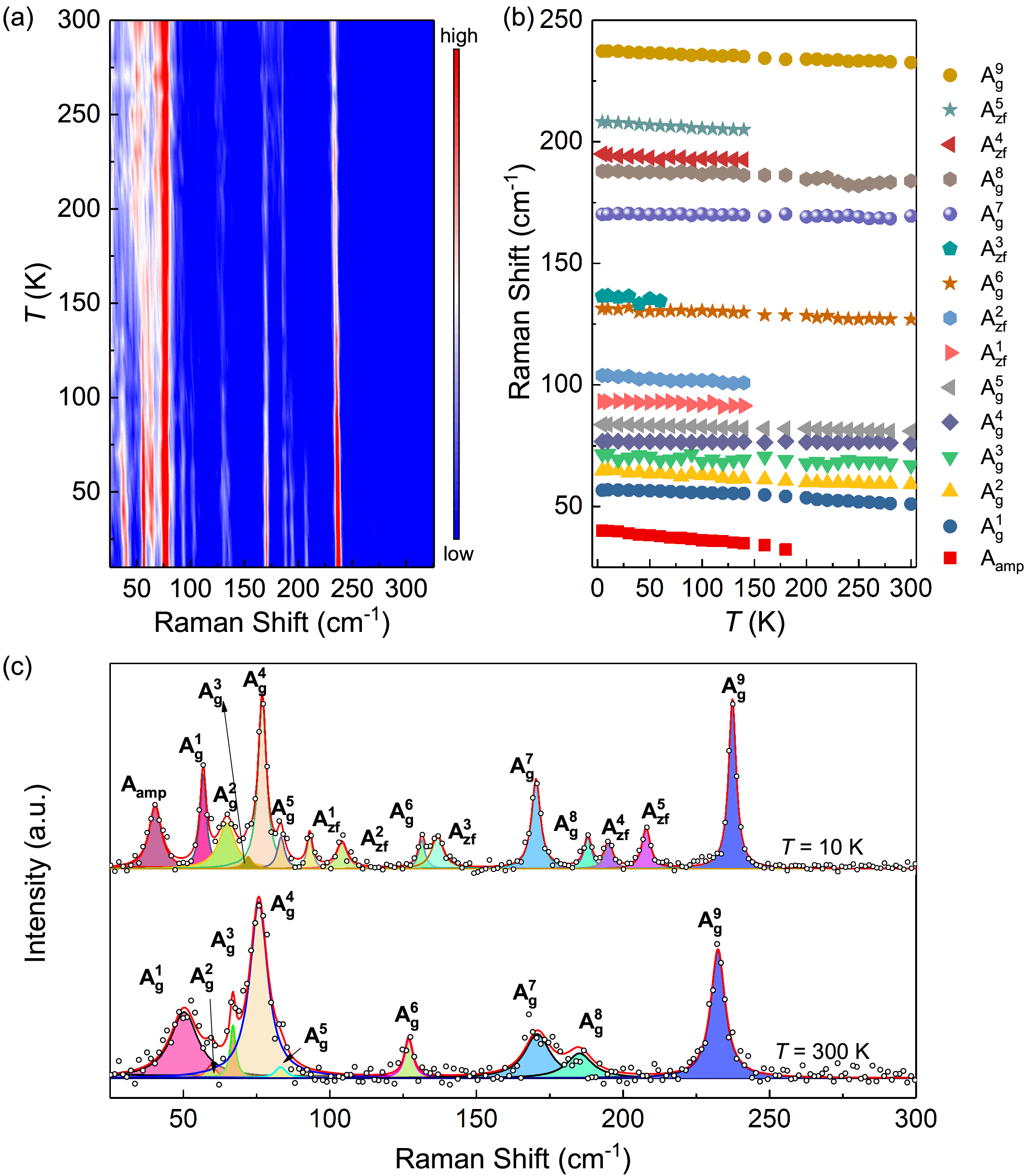}
\caption{Raman spectra of Bi$_2$Rh$_3$Se$_2$. (a)Contour map of the Raman scattering intensity. (b)Temperature dependence of Raman shift for indicated Raman modes. (c)Raman spectra of Bi$_2$Rh$_3$Se$_2$ at T = 10 and T = 300 K, respectively.}
\end{figure*}

In order to clarify the nature of the CDW phase transition at 240 K, a comprehensive study of angle- and temperature-dependant Raman scattering on Bi$_2$Rh$_3$Se$_2$ single crystals was performed to probe low-energy electronic fluctuations and phonon anomalies. Figure 2(a) shows the contour figure of temperature-dependent Raman intensity with $\theta$ = 0 for Bi$_2$Rh$_3$Se$_2$, where the angle between the polarizations of the incident light and the principal axis (\textit{a}-axis) is defined as $\theta$. In Figure 2(a), all the modes exhibit red-shift with increasing temperature due to lattice expansion, however, the modality of the frequency shift is different due to the different lattice anharmonicity of each mode.

To identify the Raman vibration modes, we performed Angle-Resolved Polarized Raman measurements at \textit{T} = 10 K. The raw experimental data for all measured Raman intensity are plotted in a contour map, as shown in Fig. S3(a)of Supplementary Materials. The frequency and linewidth of all the modes were extracted by using Lorentzian or Fano fits. When the incident light strikes along the direction of the $\textit{c}$-axis in parallel polarization configuration, the intensity of the collected Raman modes changes with the angle $\theta$ as:

\begin{equation}
\begin{split}
I\propto\lvert \vec{e}_i\cdot R \cdot\vec{e}_s\rvert =\lvert(cos\theta \;sin\theta \;0)
\left(
\begin{array}{c}
a\ b\ c\\
d\ e\ f\\
g\ h\ i\\
\end{array}\right)
\left(
\begin{array}{c}
cos\theta\\
sin\theta\\
0\\
\end{array}\right)
\rvert^2\\
=\lvert acos^2\theta+(b+d)sin\theta cos\theta+esin^2\theta\rvert^2,
\end{split}
\end{equation}

where $\vec{e}_i$ represents the unit vector of the incident laser polarization direction, $\vec{e}_s$ represents the polarization of the scattered light direction unit vector, R is the Raman tensor. The Raman tensor for $A_g$ and $B_g$ modes can be represented as:

\begin{eqnarray}
R_{A_g}=\left(
\begin{array}{c}
a\ d\ 0\\
d\ b\ 0\\
0\ 0\ c\\
\end{array}\right),\notag
R_{B_g}=\left(
\begin{array}{c}
0\ d\ 0\\
d\ 0\ 0\\
0\ 0\ 0\\
\end{array}\right)\notag
\end{eqnarray}

Then, their Raman intensities are given by

\begin{equation}
\begin{aligned}
I(A_g)=a^2cos^4\theta+b^2sin^4\theta+4d^2sin^2\theta cos^2\theta+\\2adsin2\theta cos^2\theta+2absin^2\theta cos^2\theta+2bdsin2\theta sin^2\theta,
\end{aligned}
\end{equation}
\begin{equation}
I(B_g)=0,
\end{equation}

From Eqs.(2) and (3), it can be concluded that only A$_{\rm g}$ modes with two-fold symmetry are visible under the present configurations. The angle-dependent intensity of Raman modes can be fitted by Eq. (2), the fitting results and experimental data of six selected modes with strong intensity are plotted in Fig. S3(d), represented by red solid lines and black open circles. It is notable that the new emerging modes (A$_{\rm amp}$, A$_{\rm zf}^1$ - A$_{\rm zf}^5$) also manifest two-fold symmetry [see Fig. S3(d)], implying that they are all of A-type symmetry.

At room temperature, three acoustic and thirty nine optical phonon modes are predicted according to DFPT calculations at $\Gamma$ point, where 18 modes (10A$_g$ and 8B$_g$) are Raman-active in the range of 0 - 300 cm$^{-1}$\cite{WangYS}. As shown in the lower panel of Fig. 2(c), only nine Raman-active modes were resolved (label as A$_{\rm g}^1$ - A$_{\rm g}^9$) at 300 K in our experiments, because that the intensities of other phonon modes are too weak to be distinguished. At \textit{T} = 10 K (shown in the upper panel of Fig. 2(c)), additional six modes emerge in the Raman spectra, which are labeled as A$_{\rm amp}$, A$_{\rm zf}^1$ - A$_{\rm zf}^5$, whose assignments will be discussed below. Among all modes, the A$_{\rm amp}$ mode of 39.94 cm$^{-1}$ (at \textit{T} = 10 K) shows stronger softening and appreciable broadening with increasing temperature than those of other modes, as shown in Figs. 2 (a) and 2(b), then disappearing before 240 K. All above features provide clear evidences that the A$_{\rm amp}$ here is the amplitude mode of CDW phase\cite{SugaiS}. Owing to the origin of amplitude mode, it would disappear at the CDW transition temperature due to the collapse of coherent CDW order\cite{LiuG}. But it is often heavily damped and could not be resolved before approaching the transition temperature. Moreover, the ultrafast pump-probe study on Bi$_2$Rh$_3$Se$_2$ found a signal of collective amplitude mode at $\Omega_A$ = 1.25 THz (1 THz $\sim$ 33.3 cm$^{-1}$) at low temperature\cite{LinT}, which corresponds to the same phonon energy of this A$_{\rm amp}$ mode (39.94 cm$^{-1}$) observed in our work.

\begin{figure}
\includegraphics[scale=0.5]{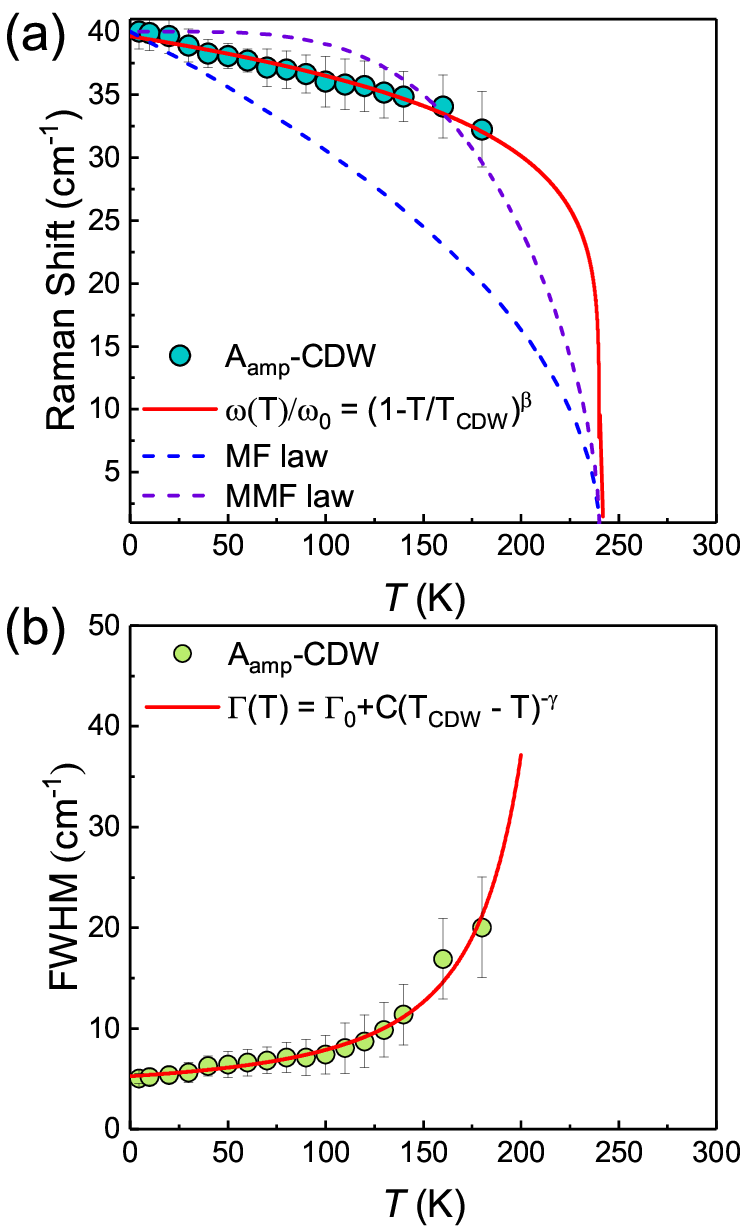}
\caption{Temperature dependence of amplitude mode. (a) and (b) show the temperature dependence of Raman shift and FWHM of A$_{\rm amp}$ mode, respectively. The solid red lines are fitting curves by using models described in the main text. The blue and purple dashed lines are fitting curves by using mean-field(MF) and modified mean-field model(MMF): $\omega(T) = \omega_0 [1-(x^4/3)]\sqrt{1-x^4}$, where $x = T/T_{CDW}$ and T$_{CDW}$ = 240 K.}
\end{figure}

The temperature evolution of amplitude mode could reflect the thermodynamic characteristics of order parameters and correlation mechanisms of the CDW state. Recently, theoretical studies have been performed by using unbiased quantum Monte Carlo simulations in the square lattice Hubbard-Holstein model, to explore the competition and interplay between CDW and AFM, as well as the coexisting SC when the electron-electron interaction is larger than a critical value\cite{OhgoeT,CostaNC,KarakuzuS}. Besides the electron-electron interaction, a temperature induced 2D Ising-type CDW phase transition would occur in the region of moderate electron-phonon coupling (EPC)\cite{CostaNC,WeberM,zhangcharge,ChenC}.

Figure. 3 shows the Raman shift and full width at half maximum(FWHM, $\Gamma$) of the amplitude mode as a function of temperature, respectively. The critical exponents of the CDW phase transition can be extracted by analysing temperature-dependent behavior of the amplitude mode\cite{ChenC}. A microscopic analysis of the nature and origin of temperature-dependent mode softening can be performed by using Rice et al.'s mean-field result\cite{RiceMJ,leeconductivity} or modified mean-field model\cite{Benfatto,WangSY} for the frequency of a CDW amplitude mode,

\begin{equation}
\omega(T) \propto \sqrt{1-T/T_{CDW}}
\end{equation}

which has been successfully applied to the analysis for the behavior of CDW soft mode in many CDW materials\cite{grassetpressure,WangSY}, as well as the temperature dependence of intensity of low-energy electronic bands for Bi$_2$Rh$_3$Se$_2$\cite{LiuZT}. Here, the CDW transition temperature $T_{CDW}$ is 240 K, extracted from the above resistivity and susceptibility data.

In Fig. 3 (a), the blue and purple dashed curves represent the fitting results of mean field theory and modified mean-field model, respectively, which dramatically deviates from our experimental data of the amplitude mode. The similar deviations from the square root mean field dependence of CDW phonon softening behavior are also observed in some two-dimensional CDW materials materials, such as GdTe$_3$,2H-NbSe$_2$,1T-TiSe$_2$,Cu$_x$TiSe$_2$\cite{chen2019raman,caopre,CuiL,BarathH}, owing to the strong CDW fluctuations\cite{JoshiJ}.

Instead, the Raman shift of amplitude mode can be fitted well by using the fitting function with a varying power $\beta$\cite{JoshiJ}, as displayed by the red line in Fig. 3(a):

\begin{equation}
\omega(T)/\omega_0=(1-T/T_{CDW})^\beta,
\end{equation}

where $\omega_0$ is the phonon frequency at $T$ $\sim$ 0 K and the fitting parameter $\beta$ is an exponent of critical behavior, \textit{T}$_{\rm CDW} = 240 K$ is extracted from the above resistivity and susceptibility data. The fit by Eq. 5 gives the values of $\omega_0$ = 40.03 cm$^{-1}$ and $\beta$ = 0.145 $\pm$ 0.032. The value of exponent $\beta$ matches the theoretical value of 1/8 for 2D Ising model\cite{stanley1971phase}.

In Fig. 3(b), temperature dependence of the full width at half maximum(FWHM, $\Gamma$) of the amplitude mode was also plotted and fitted by an empirical formula\cite{WangSY,BarathH}:

\begin{equation}
\Gamma(T)=\Gamma_0+A(T_{CDW}-T)^{-\gamma},
\end{equation}

where $\Gamma$$_0$ is the FWHM at $T$ = 0 K, A and $\gamma$ are fitting parameters, $\textit{T}$$_{\rm CDW}$ = 240 K.  The obtained parameters are $\Gamma$$_0$ = 4.34 cm$^{-1}$, A = 8.98 $\times$ 10$^5$ and $\gamma$ = 1.593 $\pm$ 0.244. This value of exponent $\gamma$ is also consistent to the theoretical value $\gamma$ = 7/4 of 2D Ising model\cite{stanley1971phase}.

Moreover, the intensity of amplitude mode becomes difficult to distinguish with increasing temperature above 180 K, before it reaches CDW transition temperature $\sim$ 240 K. At the same time, the line width of amplitude mode broadens rapidly. The temperature dependence of the amplitude mode can be attributed to the overdamped nature due to strong CDW fluctuations in ICCDW state\cite{BarathH}. At 180 K, the frequency of the amplitude mode is about 32 cm$^{-1}$, suggesting that the amplitude mode is unlikely to be associated with the ICCDW to CCDW transition, because in that case, the frequency should be near zero. This overdamping of the amplitude mode make it is impossible to observe the complete amplitude softening, but we are still able to fit the temperature- dependent amplitude mode frequency and linewidth data, because we know the exact value of CDW transition temperature.

According to the above temperature-dependent behavior of the amplitude mode, we can conclude that the phase transition at $T \sim$ 240 K can be classified into the two-dimensional Ising universality class charge transition\cite{HattaS}, suggesting that the interlayer interactions of Bi$_2$Rh$_3$Se$_2$ are negligible for the formation of electronic states. Meanwhile, due to the limit of dimensionality, the effects of quantum fluctuations would be strongly enhanced. For Ising magnetic systems, the quantum fluctuations are introduced by applying a transverse magnetic field, creating a quantum Ising model, which has been realized in one-dimensional systems\cite{wangexperimental,fauretopological,coldeaquantum}. For 2D systems, the quantum Ising model can be used to describe the quantum transition\cite{zhangcharge}. However, most of the investigations is still focused on theoretical models, more experimental researches are expected and more real systems need to be explored. Here, in Bi$_2$Rh$_3$Se$_2$, the Ising-type charge ordering is modulated by strong quantum fluctuations, which may cause quantum transition at T = 0 K and correspond to the low-temperature superconducting phase.

Besides the amplitude mode, with the emergence of the CDW state, in general the zone-folding (ZF) modes would emerge in Raman response\cite{GrunerG}. The ZF modes correspond to a normal phonon folded into the center of the Brillouin zone due to the establishment of the CDW phase. The electronic structure study of Bi$_2$Rh$_3$Se$_2$ by Angle-resolved photoemission spectroscopy (ARPES) experiments show the reconstruction of electronic structure\cite{LiuZT}. And the fact that these modes show a minute change of frequency(as shown in [Fig. 2(b)]), more consistent with the characteristics of a ZF mode\cite{SamnakayR,JoshiJ}. Thus, in addition to the amplitude mode, the new-born modes below 150 K could be labelled as ZF modes(A$_{\rm zf}^1$-A$_{\rm zf}^5$).

\begin{figure*}
\includegraphics[scale=0.5]{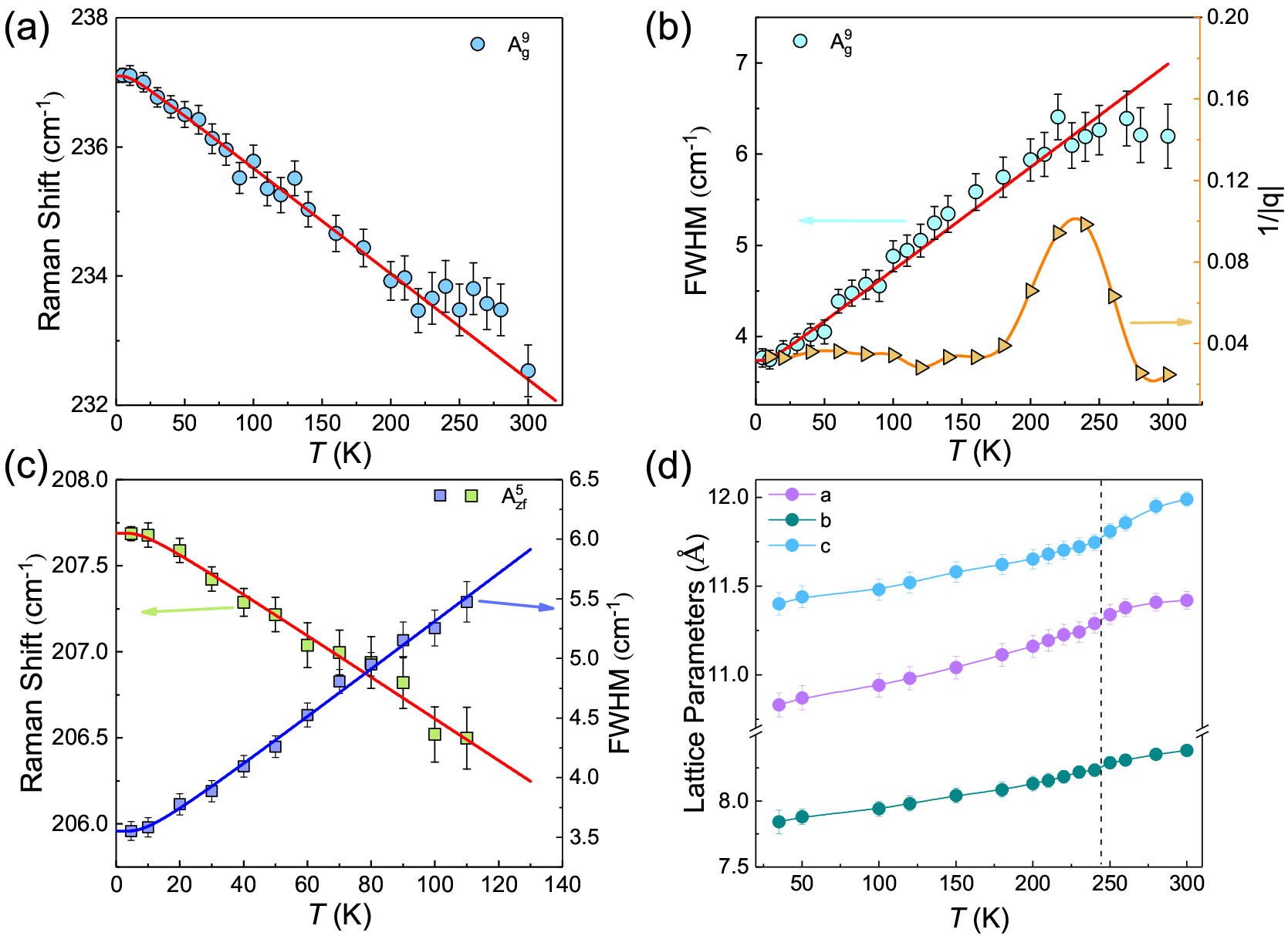}
\caption{Temperature-dependent A$_{\rm g}^9$ and zone-fold mode A$_{\rm zf}^5$.(a)-(c) Show the temperature dependence of Raman shift and FWHM of A$_{\rm g}^9$ and A$_{\rm zf}^5$ modes, respectively. The solid lines are fitting curves by using anharmonic phonon model. (d) Temperature dependence of lattice parameters a, b, and c, the solid line is a guide to the eye.}
\end{figure*}

In Raman spectra, isotopic and/or anharmonic effects, and electron-phonon interactions cause the change of linewidth. Temperature dependance of the phonon mode energy ($\omega$) and linewidth ($\Gamma$) are usually governed by phonon-phonon interaction (anharmonic effects)\cite{Lazarevic,BalkanskiM,LazarevicN,LazarevicN2}. In order to further quantitatively evaluate the lattice dynamic properties of Bi$_2$Rh$_3$Se$_2$, we fitted the temperature dependence of $\omega$ and $\Gamma$ by using the following anharmonic phonon model\cite{LazarevicN}:

\begin{equation}
\omega(T)=\omega_0-B(1+\frac{2}{e^\frac{\hbar\omega_0}{2k_BT}-1}),
\end{equation}
\begin{equation}
\Gamma(T)=\Gamma_0+C(1+\frac{2}{e^\frac{\hbar\omega_0}{2k_BT}-1}),
\end{equation}

where B and C are anharmonic constant and k$_B$ is the Boltzmann constant. In some CDW materials, the phonons may also be coupling with other fundamental excited states(i.e. electrons)\cite{LazarevicN,HuYW}, consequently an additional term $\Gamma_0$ must be included.

Two most intense modes (A$_{\rm g}^9$ and A$_{\rm zf}^5$) were selected to demonstrate the evolution of Raman modes with temperature below the CDW transition in Bi$_2$Rh$_3$Se$_2$. The corresponding fitting parameters of all the modes are given in Table S2. As shown in Fig. 4, both A$_{\rm g}^9$ and A$_{\rm zf}^5$ modes can be well fitted by the anharmonic phonon model, suggesting a stable CDW state with the robust lattice distortions below $\textit{T}$$_{\rm CDW}$\cite{WangSY}. A$_{\rm zf}^5$ mode fades away well below the CDW transition temperature due to the strong thermal fluctuations. It is worth noting that the A$_{\rm g}^9$ mode deviates from the fitting curve when it is close to the CDW transition temperature, implying the temperature-induced anharmonicity effect cannot describe the change of slope at $\textit{T}$$_{\rm CDW}$. And the A$_{\rm g}^9$ exhibits blue shift with the increasing temperature, as shown in Fig. 4(a), by which the effect of lattice expansion can be ruled out. Owing to the above temperature dependence, we can conclude that the dramatic change in A$_{\rm g}^9$  mode at 240 K comes from strong temperature-dependent electron-phonon interactions\cite{LazarevicN}.

Electron-phonon coupling is known to lead to asymmetric peak shapes, which is attributed to Breit-Wigner-Fano (BWF) resonance (or Fano resonance). As shown in Fig. S3(c), the peak of A$_{\rm g}^9$ mode shows a slight asymmetric, which can be fitted well by the Fano model \cite{FanoU}:

\begin{equation}
I=I_0\frac{[1+2(\omega-\omega_0)/q\Gamma]^2}{1+[2(\omega-\omega_0)/\Gamma]^2},
\end{equation}

where I$_0$ is the peak intensity at the peak position $\omega_0$, $\Gamma$ corresponds to the full width at half maximum (FWHM). 1/$\mid$q$\mid$ is the degree of electron-phonon coupling which describes the departure of the line shape from a symmetric Lorentzian function. The positive and negative signs of q indicate the relative position of the asymmetry, respectively. The Lorentzian shape is obtained when 1/$\mid$q$\mid$$\rightarrow$0. The 1/$\mid$q$\mid$ values before and after the phase transition are almost the same as displayed by the dark yellow line in Fig. 4(b). This means that the main contribution to the temperature dependence of the linewidth comes from the phonon-phonon(anharmonic) interactions, because the electron-phonon contribution is temperature independent(1/$\mid$q$\mid$ $\sim$ const)\cite{LazarevicN,WangYS,Gleason}. At $T \sim$ 240 K, 1/$\mid$q$\mid$ reaches the largest value, which further indicates that this phase transition is related to the electron-phonon coupling. From ARPES measurements by Liu et al.\cite{LiuZT}, a peak-dip-hump feature can be found near the Fermi momentum, which is manifested the existence of electron-phonon coupling. Therefore, by the analysis of Raman spectra, we believe that electron-phonon coupling may play an important role in the formation of CDW phase in Bi$_2$Rh$_3$Se$_2$.

Beside the electron-phonon coupling, the change of electronic structure could be a possible contribution to the change of lineshape and linewidth of Raman spectra. According to the previous literature of ARPES experiments, the change of electronic structure was observed below the CDW transition temperature $\sim$ 240 K\cite{LinT}. However, the change of density of states may occur at a lower temperature due to the gradually enhanced hybridization between replica bands and primary bands with the decreasing temperature. The Hall effect is able to provide simple indications about the electronic structure and the density of states at Fermi level. Hall resistivities at various temperatures were measured and presented in Fig. S2(a) of Supplementary Materials. It shows linear magnetic field dependence when $T \geq$ 30 K, indicating the electrical transport is dominated by the electron carriers, although the ARPES results reveal both electron and hole pockets at the Fermi level. Meanwhile, a non-linear dependence of Hall resistivity was observed at low temperatures, indicating a change of Fermi surfaces and transport properties. The low temperature Hall conductivities were fitted by two-carrier model with formula Eq. S1\cite{AliMN}, as shown in Fig. S2(b). The carrier density at different temperatures was extracted and plotted in Fig. S2(c). It exhibits a slight temperature dependence between 200 and 300 K, due to the thermal effect of lattice with changing temperature. However, there is no clear change of density of states at the CDW transition $\sim$ 240 K, implying that this transition is dominated by the electron-phonon coupling, and the contribution of electronic structure could be negligible. In fact, the reasons influencing the Fano line shape also include the density of states and the light-electron matrix element. Therefore, further efforts will be required in the future to eliminate their effects.

To understand the evolution of lattice structure around the CDW transition, the powder XRD at variation temperature from 35 K to 300 K were measured and shown in Figs. S4(a). The derived lattice parameters and volume as functions of temperature are presented in Fig. 4(d) and Fig. S4(b). No structural transition was observed with changing temperature. A distinct kink was observed at $\sim$ 240 K, corresponding to the CDW phase transition. This indicates the correlation between lattice modulation and CDW instability.

\section{\label{sec:level1}CONCLUSIONS}
To summary, by using temperature-dependent polarized Raman spectra, a second-order CDW phase transition with distinctive features near $\textit{T}$$_{\rm CDW}$ = 240 K has been explored on single crystals of Bi$_2$Rh$_3$Se$_2$. One CDW amplitude mode and five ZF modes are identified in terms of the present Raman scattering configurations, which could be fingerprinting characteristics for the CDW state. Temperature-dependent behavior of the amplitude mode indicates that the CDW state below $T \sim$ 240 K can be classified to two-dimensional Ising universality model, where the interlayer interactions could be negligible for the formation of electronic states of Bi$_2$Rh$_3$Se$_2$. Meanwhile, the simultaneous strong quantum fluctuations make Bi$_2$Rh$_3$Se$_2$ an ideal 2D system to study electronic quantum transition and the effects of quantum fluctuations on charge ordering. Furthermore, the unexpected phonon shift and coupling constant around CDW transition temperature clearly demonstrate the strong electron-phonon coupling may be the driving force of the transition. Consequently, Bi$_2$Rh$_3$Se$_2$ is a compound with multiple exotic properties, for instant CDW, SC, strong quantum fluctuations and electron-phonon coupling, which deserves further investigations.

\begin{acknowledgments}
This work was financially supported by the National Key Research and Development Program of China (Grant No.2022YFA1602603,2021YFA1600200), the National Natural Science Foundation of China (NSFC) (Grants No. 12374049, No. 12204420, No. 12204004, No. 12174395, No. U19A2093 and No. 12004004), the Natural Science Foundation of Anhui Province (No. 2308085MA16, No. 2308085QA18), the Innovation Program for Quantum Science and Technology (No. 2021ZD0302802). A portion of this work was supported by the High Magnetic Field Laboratory of Anhui Province under Contract No. AHHM-FX-2021-03.This work was supported by Systematic Fundamental Research Program Leveraging Major Scientific and Technological Infrastructure, Chinese Academy of Sciences under contract No. JZHKYPT-2021-08.Yonghui Zhou was supported by the Youth Innovation Promotion Association CAS (Grant No. 2020443), Y. X. is supported by Anhui University through the start-up project (Project No. S020318001/020).

\end{acknowledgments}
\bibliography{apssamp}

\end{document}